\documentclass[a4paper,
               keeplastbox,   
               ]{jacow}
\usepackage{pdfpages,multirow,ragged2e} %
\makeatletter%
    \ifboolexpr{bool{xetex}}
     {\renewcommand{\Gin@extensions}{.pdf,%
                        .png,.jpg,.bmp,.pict,.tif,.psd,.mac,.sga,.tga,.gif,%
                        .eps,.ps,%
                        }}{}
\makeatother

\ifboolexpr{bool{xetex} or bool{luatex}} 
 {}                                      
 {\usepackage[utf8]{inputenc}}           

\usepackage[USenglish]{babel}

\ifboolexpr{bool{jacowbiblatex}}%
 {%
  \addbibresource{jacow-test.bib}
  \addbibresource{biblatex-examples.bib}
 }{}
\begin{document}

\title{Final Physics Design of Proton Improvement Plan-II at Fermilab}

\author{Abhishek Pathak\thanks{abhishek@fnal.gov\\This manuscript has been authored by Fermi Research Alliance, LLC under Contract No. DE-AC02-07CH11359 with the US Department of Energy, Office of Science, Office of High Energy Physics.}, and Arun Saini Fermi  National Accelerator Laboratory, Batavia, USA \\ Eduard Pozdeyev,  Thomas Jefferson National Accelerator Facility, Newport, News, USA\\ 
		}
	
\maketitle

\begin{abstract}
This paper presents the final physics design of the Proton Improvement Plan-II (PIP-II) at Fermilab, focusing on the linear accelerator (Linac) and its beam transfer line. We address the challenges in longitudinal and transverse lattice design, specifically targeting collective effects, parametric resonances, and space charge nonlinearities that impact beam stability and emittance control. The strategies implemented effectively mitigate space charge complexities, resulting in significant improvements in beam quality—evidenced by reduced emittance growth, lower beam halo, decreased loss, and better energy spread management. This comprehensive study is pivotal for the PIP-II project's success, providing valuable insights and approaches for future accelerator designs, especially in managing nonlinearities and enhancing beam dynamics.
\end{abstract}

\section{INTRODUCTION}

The Proton Improvement Plan-II (PIP-II) \cite{pip2-optimization} at Fermilab features an 800-MeV superconducting linear accelerator (Linac) and a comprehensive beam transfer line\cite{cdr} to enhance proton delivery to the Booster. The superconducting Linac, composed of continuous wave (CW)-capable structures and cryomodules, operates with an average H$^-$ beam current of 2 mA and ccelerating the beam from 2.1 MeV to 0.8 GeV.

\begin{figure}[!htb]
   \centering
   \includegraphics*[width=.8\columnwidth]{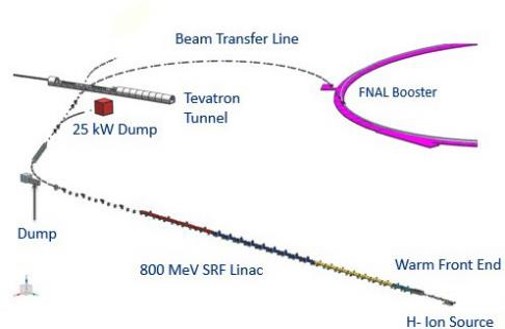}
   \caption{Schematic layout of the PIP-II Linac and associated beam transfer line: The diagram illustrates the 800 MeV SRF Linac, warm front end, H$^-$ ion source, beam transfer line, and the connection to the FNAL Booster.}
   \label{fig:layout}
\end{figure}

At such high beam intensities, space charge instabilities and parametric resonances\cite{sc} can lead to emittance growth and emittance exchange between different planes, degrading beam quality. Additionally, the presence of space charge also contributes to energy spread within the beam, which can adversely impact efficient Booster injection and increase the risk of beam loss. Therefore, it is crucial to maintain low emittance throughout the Linac and minimal momentum spread ($<2\times 10^{-4}$) at the end of the Beam Transfer Line. The Linac-to-Booster transfer line, measuring 308 meters in length, includes two arcs and a straight section, ensuring precise beam transport with minimal momentum spread. This transfer line utilizes a FODO lattice with a cell length of approximately 11.8 meters to maintain dispersion and beta-function alignment with the SC Linac and Booster. A schematic representing the SC Linac and the BTL section of the PIP-II is shown in Figure~\ref{fig:layout}.

The present design supports potential future upgrades to increase the Linac energy to 1.2 GeV, highlighting its flexibility for further enhancements in proton acceleration and beam delivery efficiency.

This paper presents a re-optimized physics design of the PIP-II superconducting Linac that eliminates all parametric resonances and provides a design that leads to a negligible transverse emittance growth of 4\% and an insignificant emittance growth of 3\% in the longitudinal plane. This optimization was performed with a systematic implementation of the design philosophies discussed in Section 1. This paper also discusses the implementation of a debuncher cavity in cell-3 of the beam transfer line to reduce the space charge-dominated energy spread growth along the beam transfer line, allowing for efficient Booster injection. 

\section{SC Linac Design Optimization Approach}

Our approach to optimizing the superconducting section of the PIP-II Linac is grounded in a well-defined design philosophy. This systematic optimization incorporates the following steps:

\begin{itemize}
    \item \textbf{Adiabatic Control of Phase Advance:} Maintain adiabatic variation of phase advance per unit length ($k\zeta_0$) to avoid large amplitude oscillations and beam size growth.
    \item \textbf{Structure Phase Advance Variation:} Keep zero current phase advance per period ($\sigma\zeta_0$) below 90° to prevent envelope instabilities.
    \item \textbf{Transverse Focusing Dynamics:} Account for RF cavity electromagnetic field contributions to $k^2\zeta_0(s, \gamma)$ that affect betatron oscillations, and prevent parametric resonance by ensuring $\sigma t_0$ does not equal an integer multiple of $\sigma z_0/2$.
    \item \textbf{Mitigating Space Charge Effects:} Position lattice footprints in resonance-free zones on the Hofmann diagram to prevent emittance exchange; maintain tune depression above 0.5.
    \item \textbf{Optimal Beam Matching Across Lattice Sections:} Implement proper beam matching between lattice sections to reduce beam size mismatch and halo formation.
\end{itemize}

\begin{figure}[!htb]
   \centering
   \includegraphics*[width=.8\columnwidth]{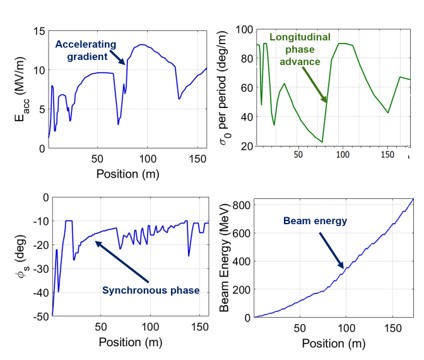}
   \caption{Lattice parameters for the superconducting section of the PIP-II Linac after re-optimization: (a) Accelerating gradient (E$_{acc}$) along the Linac, showing the variation in gradient with position, (b) Longitudinal phase advance per period ($\sigma_0$) as a function of position, demonstrating control over phase advance to prevent instabilities, (c) Synchronous phase ($\phi_s$) along the Linac, indicating phase stability during acceleration, and (d) Beam energy profile along the Linac, depicting the energy gain from 2.1 MeV to 800 MeV.}
   \label{fig:lattice_parameters}
\end{figure}

By adopting this comprehensive design strategy, we have re-optimized the physics design of the superconducting section of the PIP-II Linac. This re-optimization eliminates parametric resonances and provides a design that achieves negligible transverse emittance growth of 4\% and insignificant emittance growth of 3\% in the longitudinal plane. The optimized lattice parameters are shown in Figure~\ref{fig:lattice_parameters}.

\begin{figure}[!htb]
   \centering
   \includegraphics*[width=.8\columnwidth]{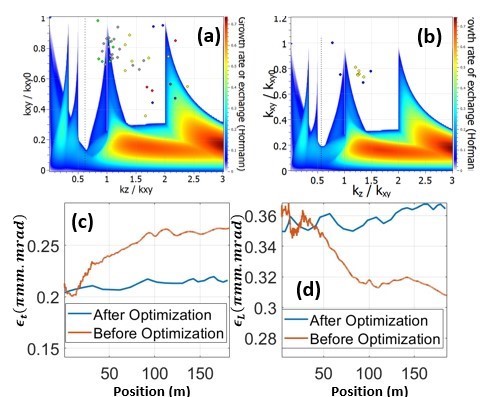}
   \caption{Impact of optimization on beam stability and emittance in the PIP-II Linac: (a) and (b) show the Hofmann's chart before and after optimization, respectively, highlighting the elimination of parametric resonances. (c) and (d) display the transverse and longitudinal normalized rms emittance before and after optimization along the superconducting section of the Linac, showing elimination of emittance exchange.}
   \label{fig:optimization_impact}
\end{figure}

After implementation of the lattice shown in Figure~\ref{fig:lattice_parameters}, 6D beam matching and a multi-variable global optimization were performed using machine learning algorithms such as genetic algorithms, Convolutional Neural Networks (CNNs), and Random Forests to further improve the beam emittance and to compensate for transverse inter-planar coupling attributed to asymmetric quadrupolar kicks\cite{asym-1,asym-3} in the spoke cavities. Figure~\ref{fig:optimization_impact} demonstrates the effectiveness of our systematic approach accompanied by machine learning algorithms, showing the absence of parametric resonances in the stability chart and therefore eliminating emittance exchange between the longitudinal and transverse planes, and reducing transverse and longitudinal beam emittance throughout the Linac.

\begin{figure}[!htb]
   \centering
   \includegraphics*[width=.8\columnwidth]{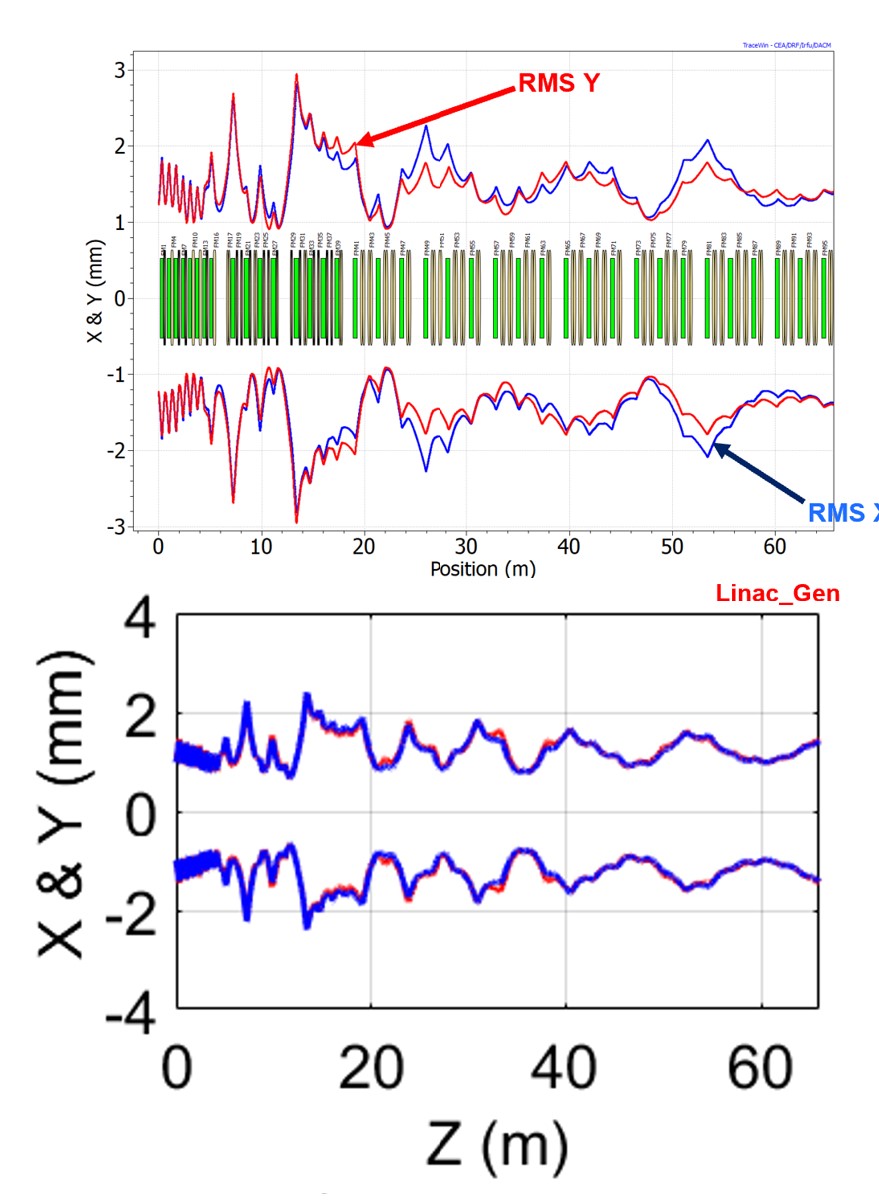}
   \caption{RMS beam sizes before and after optimization of the solenoid field. The top graph shows the RMS sizes before optimization, while the bottom graph shows the RMS sizes after optimization. This optimization reduced the difference between the x and y RMS sizes from 20\% to 5\%.}
   \label{fig:solenoid_optimization}
\end{figure}

As it was mentioned, the asymmetry of the spoke cavities induces a quadrupolar transverse momentum kick to the beam, leading to an elliptical transverse particle density distribution. When such a beam enters a solenoid focusing channel, the elliptical distribution rotates and causes transverse inter-planar coupling. Although this effect is dominant at lower energy, the transverse defocusing of the RF cavities reduces quadratically with beam relativistic $\beta$. Taking advantage of this fact and using reinforcement learning techniques, we tuned the solenoids' field in a way that minimizes the difference between the x and y rms sizes while maintaining a lower value of beam emittance. Figure~\ref{fig:solenoid_optimization} demonstrates the rms beam sizes before and after optimization of the solenoid field. This optimization process reduces the splitting between the x and y rms sizes from 20\% to 5\%.

\section{Longitudinal Dynamics in Beam Transfer Line}

\begin{figure}[!htb]
   \centering
   \includegraphics*[width=.8\columnwidth]{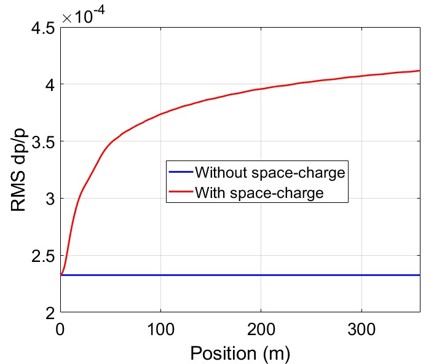}
   \caption{Evolution of the RMS momentum spread (dp/p) along the beam transfer line, shown with (red line) and without (blue line) space charge effects.}
   \label{fig:momentum_spread}
\end{figure}

The longitudinal phase-space at the exit of the Linac demonstrates a momentum spread of $2\times 10^{-4}$. As the beam propagates towards the Booster through the beam transfer line, in the absence of longitudinal focusing, space charge causes the beam momentum spread to increase, and at the end of the BTL, the momentum spread increases to an rms value of $4.2\times 10^{-4}$, which is more than double the required momentum spread of $2\times 10^{-4}$ for efficient Booster injection. The evolution of the momentum spread with and without space charge is shown in Figure~\ref{fig:momentum_spread}.

\begin{figure}[!h]
   \centering
   \includegraphics*[width=1\columnwidth]{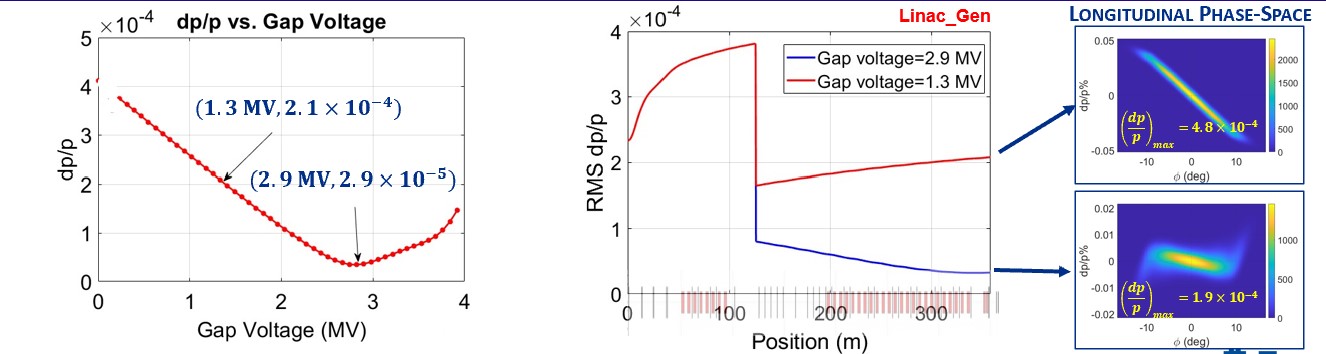}
   \caption{(Left) Variation of RMS momentum spread ($dp/p$) with gap voltage, showing required $\frac{dp}{p}$ of $2 \times 10^{-4}$ at 1.3 MV and minimum $\frac{dp}{p}$ of $2.9 \times 10^{-5}$ at 2.9 MV. (Middle) Evolution of the RMS momentum spread ($dp/p$) along the beam transfer line at 1.3 MV (red line) and 2.9 MV (blue line). (Right) Corresponding output longitudinal phase space distribution at 1.3 MV and 2.9 MV.}
   \label{fig:momentum_spread_variation}
\end{figure}

As one of the reasons for the increase in the momentum spread is the absence of longitudinal focusing, we decided to study the effectiveness of a debuncher cavity in the BTL. Given the limitations in terms of space availability, we decided to place the debuncher in the third cell of the BTL. Once the location of the cavity was decided, extensive studies were performed to investigate the choice of operating frequency, gap voltage, and the effect of beam and cavity misalignment on the beam quality and beam transmission. Finally, a cavity operating at a frequency of 650 MHz was adopted based on a compromise between the increase in cavity length and the required gap voltage at lower frequencies, and the choice of technology (superconducting or room temperature). Figure~\ref{fig:momentum_spread_variation} demonstrates the variation of momentum spread with an increase in the gap voltage, achieving the required $\frac{dp}{p}$ of $2 \times 10^{-4}$ at a gap voltage of 1.3 MV. With further increase of the gap voltage, we can further reduce the momentum spread to achieve the minimum value of $2.9 \times 10^{-5}$ at a gap voltage of 2.9 MV.

Figure~\ref{fig:momentum_spread_variation} also illustrates the rms $\frac{dp}{p}$ variation along the BTL at 1.3 MV and 2.9 MV with the corresponding output longitudinal phase space distribution. This study suggests that the use of such a debuncher cavity can compensate for the space charge dominated momentum spread increase in the BTL and facilitate efficient Booster injection without causing beam loss.

\section{Conclusion}

The final physics design of the Proton Improvement Plan-II (PIP-II) at Fermilab presents an optimized solution for the superconducting Linac and its beam transfer line. Utilizing strategies such as adiabatic control of phase advance, structure phase advance variation, and transverse focusing dynamics, we successfully mitigated space charge effects and parametric resonances, resulting in negligible transverse emittance growth of 4\% and insignificant longitudinal emittance growth of 3\%.

Advanced machine learning algorithms, including genetic algorithms, Convolutional Neural Networks (CNNs), and Random Forests, further improved beam emittance and minimized transverse inter-planar coupling induced by the asymmetry of the spoke cavities. This comprehensive optimization demonstrated a significant reduction in transverse and longitudinal beam emittance throughout the Linac.

The integration of a 650 MHz debuncher cavity in the third cell of the beam transfer line effectively reduced the momentum spread from $4.2 \times 10^{-4}$ to the required $2 \times 10^{-4}$, ensuring efficient Booster injection and minimal beam loss.

The results of this study are pivotal for the success of the PIP-II project, providing a robust framework for managing nonlinearities and enhancing beam dynamics. The optimized lattice parameters and beam characteristics shown in Figures 2 through 6 highlight the efficacy of our design and optimization approaches, contributing significantly to the advancement of high-intensity proton accelerator technologies.

%
%
\ifboolexpr{bool{jacowbiblatex}}%
    {\printbibliography}%

\begin{thebibliography}{9} 
   
\bibitem{pip2-optimization}
Pathak, A., \& Pozdeyev, E. \textit{"Optimization of the Superconducting Linac for Proton Improvement Plan-II (PIP-II)."} 5th North American Particle Accel. Conf. NAPAC2022, Albuquerque, NM, USA (2022). doi:10.18429/JACoW-NAPAC2022-MOPA36.



    
    \bibitem{cdr}
Lebedev, Valeri. \textit{"The PIP-II reference design report."} Technical Report No. FERMILAB-DESIGN-2015-011607162 (2015).
 
 
 
 
 
 \bibitem{sc}
 Hofmann, Ingo. \textit{"Stability of anisotropic beams with space charge."} Physical Review E 57, no. 4 (1998): 4713. doi: {10.1103/PhysRevE.57.4713}

    
 \bibitem{asym-1}
Berrutti, P., M. H. Awida, B. Shteynas, I. V. Gonin, N. Solyak, V. P. Yakovlev, and A. Saini. \textit{"Effects of the RF Field Asymmetry in SC Cavities of the Project X."} WEPPC047, IPAC12 proceedings (2012).

\bibitem{asym-3}
Berrutti, Paolo, Timergali Khabiboulline, Valeri Lebedev, and Vyacheslav Yakovlev. \textit{"Transverse field perturbation for PIP-II SRF cavities."} In 6th International Accelerator Conference (IPAC2015). 2015.

 
    \end{thebibliography}
    {%
    

}
\end{document}